\documentclass[apjl]{emulateapj}
\usepackage{amsmath,amssymb,graphicx,soul,color}
\usepackage{epstopdf}
\usepackage{subfigure}
\usepackage{hyperref}
\usepackage{soul}

\usepackage{natbib}
\slugcomment{Submitted to ApJ Letters}

\begin{document}

\title{GRB 090926A and Bright Late-time \textit{Fermi} LAT GRB Afterglows}


\author{C. A. Swenson\altaffilmark{1}, A. Maxham\altaffilmark{2}, P. W. A. Roming\altaffilmark{1,3}, P. Schady\altaffilmark{4}, L. Vetere\altaffilmark{1}, B. Zhang\altaffilmark{2}, B. B. Zhang\altaffilmark{2}, S. T. Holland\altaffilmark{5,6}, J. A. Kennea\altaffilmark{1}, N. P. M. Kuin\altaffilmark{7}, S. R. Oates\altaffilmark{7}, K. L. Page\altaffilmark{8}, M. De Pasquale\altaffilmark{7}}

\email{cswenson@astro.psu.edu}
\altaffiltext{1} {Pennsylvania State Univ., 525 Davey Lab, University Park, PA 16802, USA}
\altaffiltext{2}{Univ. of Nevada, Las Vegas, Box 454002, 4505 Maryland Parkway, Las Vegas, NV 89154, USA}
\altaffiltext{3}{Southwest Research Institute, 6220 Culabra Road, San Antonio, TX 78238, USA}
\altaffiltext{4}{Max-Planck-Institut f\"{u}r extraterrestrische Physik, Garching, Germany}
\altaffiltext{5}{Universities Space Research Association, 10227 Wincopin Circle, Suite 500, Columbia, MD 21044, USA}
\altaffiltext{6}{Center for Research \& Exploration in Space Science \& Technology, Code 668.8, Greenbelt, MD 20771, USA}
\altaffiltext{7}{The UCL Mullard Space Science Laboratory, Holmbury St Mary, Surrey, RH5 6NT, UK}
\altaffiltext{8}{Department of Physics and Astronomy, University of Leicester, University Road, Leicester LE1 7RH, UK}

\shortauthors{Swenson, C. A. et al. 2010}

\begin{abstract}
GRB 090926A was detected by both the GBM and LAT instruments on-board the \textit{Fermi} Gamma-Ray Space Telescope.  \textit{Swift} follow-up observations began $\sim$13 hours after the initial trigger.  The optical afterglow was detected for nearly 23 days post trigger, placing it in the long lived category.  The afterglow is of particular interest due to its brightness at late times, as well as the presence of optical flares at T0+10$^5$ s and later, which may indicate late-time central engine activity.  The LAT has detected a total of 16 GRBs; 9 of these bursts, including GRB 090926A, also have been observed by \textit{Swift}.  Of the 9 \textit{Swift} observed LAT bursts, 6 were detected by UVOT, with 5 of the bursts having bright, long-lived optical afterglows.  In comparison, \textit{Swift} has been operating for 5 years and has detected nearly 500 bursts, but has only seen $\sim$30\% of bursts with optical afterglows that live longer than 10$^5$ s.  We have calculated the predicted gamma-ray fluence, as would have been seen by the BAT on-board \textit{Swift}, of the LAT bursts to determine whether this high percentage of long-lived optical afterglows is unique, when compared to BAT-triggered bursts.  We find that, with the exception of the short burst GRB 090510A, the predicted BAT fluences indicate the LAT bursts are more energetic than 88\% of all Swift bursts, and also have brighter than average X-ray and optical afterglows.
\end{abstract}


\section{Introduction}

The \textit{Fermi} Gamma-ray Space Telescope has opened a new era of gamma-ray burst (GRB) observations.  With the on-board Gamma-ray Burst Monitor (GBM) and Large Area Telescope (LAT) instruments (Atwood et al. 2009; Meegan et al. 2009), the GRB prompt emission can be probed at higher energies than ever before.  Used in conjunction with \textit{Swift} (Gehrels et. al. 2004), GRB afterglows can be studied across a nearly continuous band from GeV energies to optical wavelengths.  As of April 1, 2010 the \textit{Fermi} GBM has triggered on nearly 450 GRBs, 16 of which have also been seen by the LAT.  Of the 16 LAT-detected GRBs, one was simultaneously localized by the \textit{Swift} Burst Alert Telescope (BAT; Barthelmy et al. 2005), and 8 others had \textit{Swift} follow-up observations at late times.  The \textit{Swift} X-ray Telescope (XRT; Burrows et al. 2005a) detected the X-ray afterglow from 7 of the 9 LAT bursts; 6 of the 7 with X-ray afterglows detected by the \textit{Swift} UV/Optical Telescope (UVOT; Roming et al. 2005).  All but one of the UVOT afterglows stand out due to their brightness and length of detectability.

During the first 5 years of operation, \textit{Swift} detected nearly 500 GRBs, but only a small percentage ($\lesssim$30\%) had bright, long-lived optical afterglows that extended beyond 10$^5$ s.  Comparing the \textit{Swift} GRBs to the LAT GRBs, two-thirds of the LAT bursts with follow-up observations have optical afterglows that rival the brightest and longest lived of the \textit{Swift} sample.  Such a high percentage raises the question as to whether the LAT bursts differ from the \textit{Swift} sample.  Two possibilities are that the LAT is observing GRBs that exhibit extended energy injection, resulting in bright optical afterglows at late times, or the LAT bursts could simply be brighter, at all wavelengths, than the `average' BAT-triggered burst allowing for later detections of the afterglow (cf. Gehrels et~al.).

GRB 090926A is a LAT-detected burst with a bright, long-lived UVOT afterglow.  In this paper, we present the multiwavelength study of GRB 090926A, examining the X-ray and UV/optical wavelengths as observed by \textit{Swift}.  In an attempt to understand the high percentage of LAT-detected bursts with optical afterglows, we also use the \textit{Fermi} observations of the prompt emission to calculate the expected fluence as would have been observed by the BAT.  We perform this same calculation for the 6 other LAT bursts also detected by XRT and compare them to a sample of BAT-triggered bursts.

We use the power-law representation of flux density, $f{_\nu}(t) \propto t^{\alpha} \nu^{\beta}$, where $\alpha$ and $\beta$ are the temporal and spectral indices, respectively.  Errors are reported at 1$\sigma$, unless otherwise specified.

\section{Observations and Data Reduction}

\subsection{Fermi data}

\begin{figure}
\includegraphics[scale=0.79]{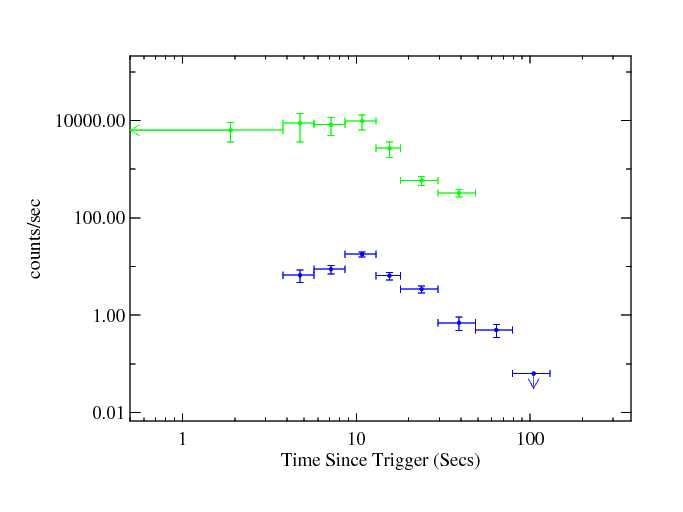}
\caption{$\textit{Fermi}$ GBM (upper) and LAT (lower) light curves.}
\label{fig:LATGBMlight curve}
\end{figure}

At 04:20:26.99 UT on 2009 September 26, the GBM triggered on GRB 090926A (Bissaldi 2009).  The GBM light curve, Fig.~\ref{fig:LATGBMlight curve}, consisted of a single pulse with T$_{90}$ of 20$\pm$2 s (8-1000 keV).  The time-averaged, combined GBM/LAT spectrum from T0 to T0+20.7 s, where T0 is the trigger time, is best fit by a Band function (Band et al. 1993), with $E_{peak}$ = 268$\pm$4 keV, $\alpha$ = -0.693$\pm0.009$ and $\beta$ = -2.342$\pm 0.011$.  The fluence (10 keV - 10 GeV) during this interval is  (2.47$\pm$0.03)$\times10^{-4}$ ergs cm$^{-2}$, bright enough to result in a \textit{Fermi} repointing.  In the first 300 s, LAT observed 150 and 20 photons above 100 MeV and 1 GeV, respectively.  Possible extended emission continued out to a few kilo-seconds.  The highest energy photon, 19.6 GeV, was observed 26 s after the trigger.  The LAT light curve, Fig.~\ref{fig:LATGBMlight curve}, is fit by a power-law of $\alpha$ = -2.17$\pm0.14$.  We fit the LAT spectrum, from 100 - 1000 s, with a power-law of $\beta = -1.26^{+0.24}_{-0.22}$.

\subsection{XRT data}

\begin{figure}
\includegraphics[scale=0.67]{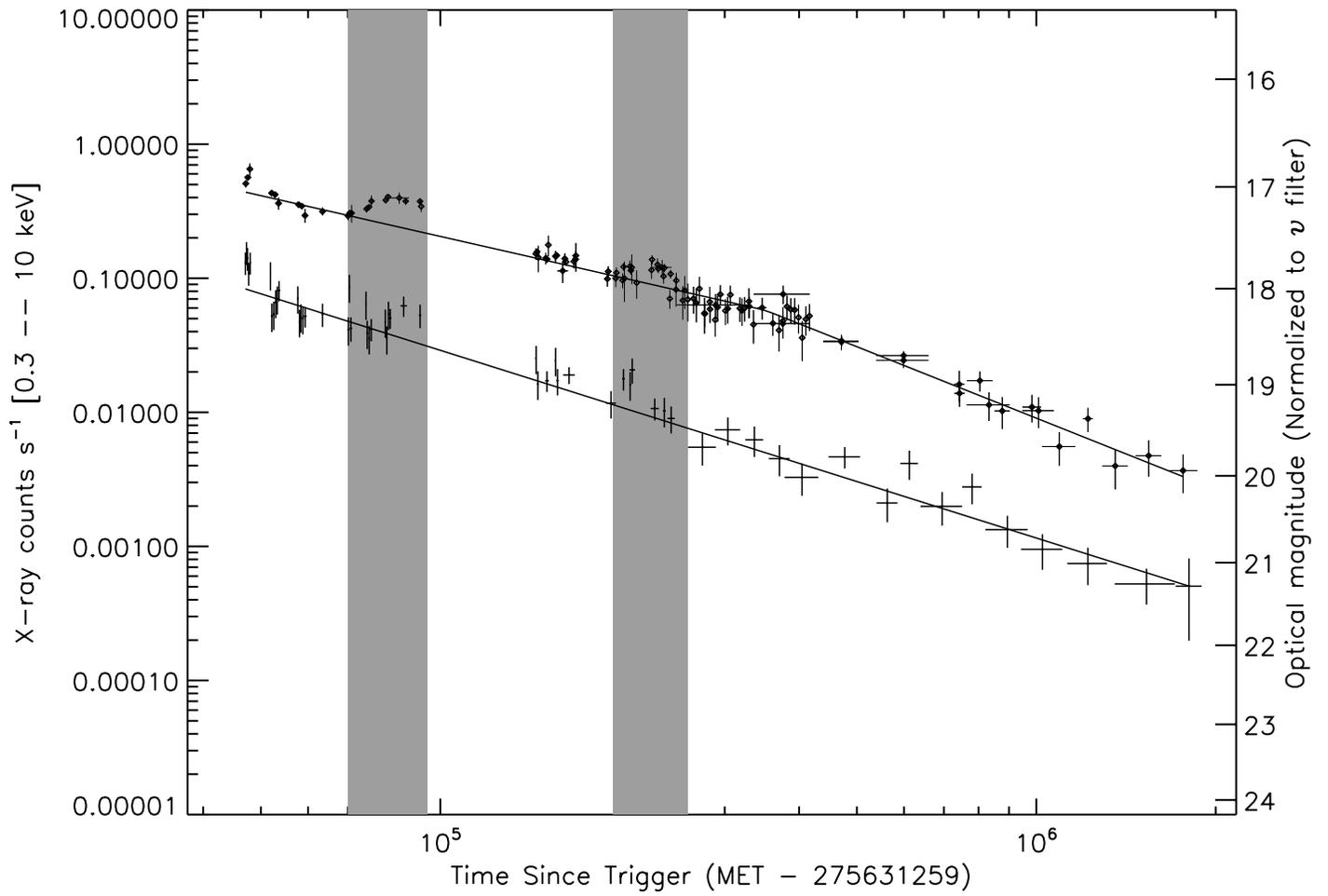}
\caption{Light curves for the XRT (bottom) and UVOT (top).  Shaded regions indicate periods of flaring.}
\label{fig:lightcurves}
\end{figure}

XRT began observing GRB 090926A $\sim$46.6 ks after the \textit{Fermi} trigger, in Photon Counting (PC) mode. The light curve, Fig.~\ref{fig:lightcurves} (taken from the XRT light curve repository; Evans et~al. 2007, Evans et~al. 2009), shows a decaying behavior with some evidence of variability, and is fit with a single power-law, decaying with $\alpha$ = -1.40$\pm$0.05 (90\% confidence level).
The average spectrum from 46.6 ks -- 149 ks is best fit by an absorbed power-law model with $\beta = -1.6^{+0.3}_{-0.2}$ and an absorption column density of $1.0^{+0.5}_{-0.3}\times10^{21}$ cm$^{-2}$ in excess of the Galactic value of 2.7$\times10^{20}$ cm$^{-2}$ (Kalberla et al. 2005). The counts to observed flux conversion factor deduced from this spectrum is 3.5$\times10^{-11}$ ergs cm$^{-2}$ count$^{-1}$. The average observed (unabsorbed) fluxes are 1.3(1.9)$\times10^{-12}$ ergs cm$^{-2}$ s$^{-1}$.

\subsection{UVOT data}

UVOT began settled observations of GRB 090926A at T0+$\sim$47 ks, and the optical afterglow was immediately detected (Gronwall et al. 2009).  The resulting optical afterglow light curve is shown in Fig.~\ref{fig:lightcurves}.  Removing these flares, the underlying optical light curve is well fit ($\chi^{2}_{red}$ = 0.92/82 d.o.f.) by a broken powerlaw.  The best fit parameters are:  $\alpha_{Opt,1} = -1.01^{+0.07}_{-0.03}$, $t_{break} = 351^{+70.2}_{-141.9}$ ks, $\alpha_{Opt,2} = -1.77^{+0.21}_{-0.26}$.  X-shooter, mounted on the Very Large Telescope UT2, found a spectroscopic redshift of $z$ = 2.1062 (Malesani et~al. 2009).

\subsection{Flaring activity}

The variability in the X-ray is not statistically strong (peaking $\sim$2.8$\sigma$ above the underlying fit) but is temporally coincident with stronger flaring in the UVOT.  The first flare, at $\sim$70 ks -- 95ks, is well defined in the UVOT lightcurve with $\Delta t$/$t$ $\approx$ 0.35, but is only seen in the X-ray as minor variability, with individual points varying from the underlying fit.  The second flare, at 195 ks -- 260 ks, is better defined in the X-ray (though only peaking at $\sim$1$\sigma$) but is matched by a similarly shaped, stronger feature in the UVOT ($\Delta t$/$t$ $\approx$ 0.28).  Due to an observing gap, we may not have observed the peak of the UVOT feature, but it appears to lag the peak of the X-ray feature by at least 6 ks, which is consistent with lower energy emission from flares laging the higher energy (Margutti 2010).

\section{Discussion}

GRB 090926A was a long burst with more than 20 photons in the GeV range, that was also easily detected by the \textit{Swift} XRT and UVOT nearly 13 hrs after the initial trigger and has late time flares in the UVOT afterglow.  The overall brightness and behavior of the optical afterglow are more reminiscent of afterglows observed immediately after the trigger, as opposed to observations starting at 47 ks after the trigger (Oates et~al. 2009; Roming et~al. 2009; Roming et~al. 2010, in preparation).  The late time light curve could be due to late time energy injection, supported by the presence of flares in the light curve, or could be a LAT selection effect.  We explore both of these possibilities.

\subsection{GRB 090926A late time flares}

X-ray flares at late times have been attributed to two different sources (Wu et~al. 2005):  central engine powered internal emission, or features of the external shock.  There is evidence suggesting that the GRB prompt emission and X-ray flares originate from similar physical processes (see Burrows et~al. 2005b; Zhang et~al. 2006; Chincarini et~al. 2007; Krimm et al. 2007), including a lower energy budget and `spiky' flares more like those actually seen in X-ray light curves.  If the central engine is the source of GRB flares, the X-ray flare spectrum should be similar to that of the prompt spectrum.  In the case of GRB 090926A, the prompt emission was seen to have a Band-function spectrum.  Assuming the optical behaves similarly to the X-ray and that the flares are caused by central engine activity, we would expect a Band-function spectrum during the flares.  A Band-function spectrum is not observed during the X-ray variability or optical flares.  The flares are both well fit by a power-law, with no indication of a break in the spectrum or sign of spectral evolution in the X-ray.  It should be stated, however, that the statistics of the X-ray light curve are low enough that detecting a Band spectrum may not be possible, even if it exists.  Combining the poor statistics with the dominate underlying continuum, it is not surprising that a power-law is the best fit.  We also find no evidence of change in the spectral shape after creating a spectral energy distribution using optical/UV photometry before and during the first flare.

A non Band-like spectrum for the flares does not expressly prohibit central engine activity from being the source of the flares, but it does allow for alternate explanations.  Code for modeling X-ray flares in GRBs developed by Maxham and Zhang (2009) can produce optical flares through the collision of low energy shells or wide shells.  If the two flares are indeed due to internal shocks, then this code can put constraints on the time of ejection and maximum energy (Lorentz factor) of
the matter shells that could produce such flares.  Since ejection time in the GRB rest frame is highly correlated to the collision time of shells in the observer frame, this means that the central engine is active around 70 ks and 197 ks.
Using the prompt emission fluence to constrain the total energy contained in the blastwave, the internal shock model requires that Lorentz factors of the shells causing flares must be less than the Lorentz factor of the blastwave when the shells are ejected. Fast moving shells will simply collide onto the blastwave giving small, undetectable glitches, whereas slow moving shells will be allowed to collide internally, releasing the energy required to detect a flare.  Specifically, we find maximum Lorentz factors of 8.2 $(\frac{E_{52.3}}{n})^{1/8}$ and 5.5 $(\frac{E_{52.3}}{n})^{1/8}$  for the first
and second flare, respectively and in terms of the energy in the prompt emission in units of $10^{52.3}$ ergs
and number density of the ambient medium.

Collisions between these relatively low energy shells are expected to be seen in the lower energy UV/optical bands.  In the synchrotron emission model, $E_p =2 \Gamma \gamma_e2 \frac{\hbar e B}{m_e c} \propto L^{1/2}$  for electrons moving with a bulk Lorentz factor $\Gamma$ with typical energy $\gamma_e m_e c^2$, since the comoving magnetic field $B \propto L^{1/2}$ (Zhang \& M{\'e}sz{\'a}ros, 2002).  This is consistent with the empirical Yonetoku relation $E_p \propto L_{iso}^{1/2}$ (Yonetoku et al. 2004) for prompt GRB emission. Applying this relation to the two flares, one predicts $E_p$ of 0.8 and 0.5 eV for each flare, respectively. This is consistent with the observation that both flares are more prominent in the optical band than in the X-ray band. Finding $E_p$ using the Amati relation, $E_p \propto E_{iso}^{1/2}$, (Amati et al. 2002) gives $E_p$ values for both flares around 1 keV, which are inconsistent with the observation. Unlike for individual burst pulses (whose durations do not vary significantly), which seem to follow an Amati relation (Krimm, et al. 2009), the Yonetoku relation may be more relevant for flares because it is consistent with the more generic synchrotron emission physics.  Since the duration of a flare depends on the epoch of the flare (the time it is seen), the Amati relation is not expected to hold.

\subsection{Are LAT bursts brighter than average?}

Despite its remarkably bright, late detection, GRB 090926 is not the first optical counterpart to be found at such late times.  Since the launch of the \textit{Fermi} satellite, \textit{Swift} has performed follow-up observations on 8 GBM triggered bursts with LAT detections:  GRBs 080916C, 081024B, 090217, 090323, 090328, 090902B, 090926A, and 091003, all of which are long GRBs.    None of these bursts were observed before $\sim$39 ks.  Although \textit{Swift} observations were performed as soon as possible, the error circle of the GBM (typical error radius of a few degrees) is too large to be effectively observed by \textit{Swift}, and the more precise LAT position was required to better constrain the error radius before observations could take place.  Despite these delays, an X-ray counterpart was discovered by XRT for 6 of the 8 bursts with follow-up observations. UVOT detected an optical afterglow associated with 5 of the X-ray counterparts.  In addition to these follow-up observations, the short GRB 090510A was a coincident trigger between GBM/LAT and BAT, raising the total number of \textit{Swift} observed LAT bursts to 9.  GRB 090510A had both an X-ray and UV/optical counterpart.

The high percentage of LAT-detected bursts with optical afterglows, when compared to the sample of \textit{Swift} triggered bursts, raises questions about the nature of the bursts themselves.  Is the LAT instrument preferentially sensitive to bursts that are brighter overall, resulting in a higher probability of detecting a bright, long-lived optical counterpart, or are the bursts themselves different, with a late time brightening causing the optical afterglows?

To investigate the former possibility, we calculated the fluence that would have been observed by the BAT for the  bursts that were triggered by \textit{Fermi}/LAT and later detected by XRT.  Because we are assuming, for the purpose of this test, that the spectrum is brighter at all wavelengths, a bright LAT burst corresponds to a bright GBM burst.  Under this assumption, we use the GBM spectral parameters provided by  Ghisellini et~al. (2010) to predict what would have been seen by the BAT over the 15-150 keV range.  We check our results and estimate our error by comparing the predicted and observed fluence for the simultaneously observed \textit{Fermi}/\textit{Swift} GRB 090510A.  The GBM spectral parameters, as well as the predicted BAT fluence between 15-150 keV are shown in Table~\ref{tab:ParamsFluence}.

We limit our error in the calculation of the expected BAT fluence to the error introduced from the GBM parameters.  Comparing the T$_{90}$ of GRB 090510A as observed by the GBM and BAT (1 s and 0.3 s, respectively), we realize that a certain amount of error will be introduced into the expected BAT fluence due to differences that would exist in the observed T$_{90}$ between the two instruments.  In the case of the long bursts, this error is negligible in comparison to the GBM parameter errors.  Because GRB 090510A is a short burst, a small difference in $T_{90}$ results in a proportionally larger error than a difference in a few seconds for longer bursts.  However, our calculated value of the fluence for GRB 090510A differs by less than a factor of two from the BAT observed value.

\begin{figure}
\includegraphics[scale=0.67]{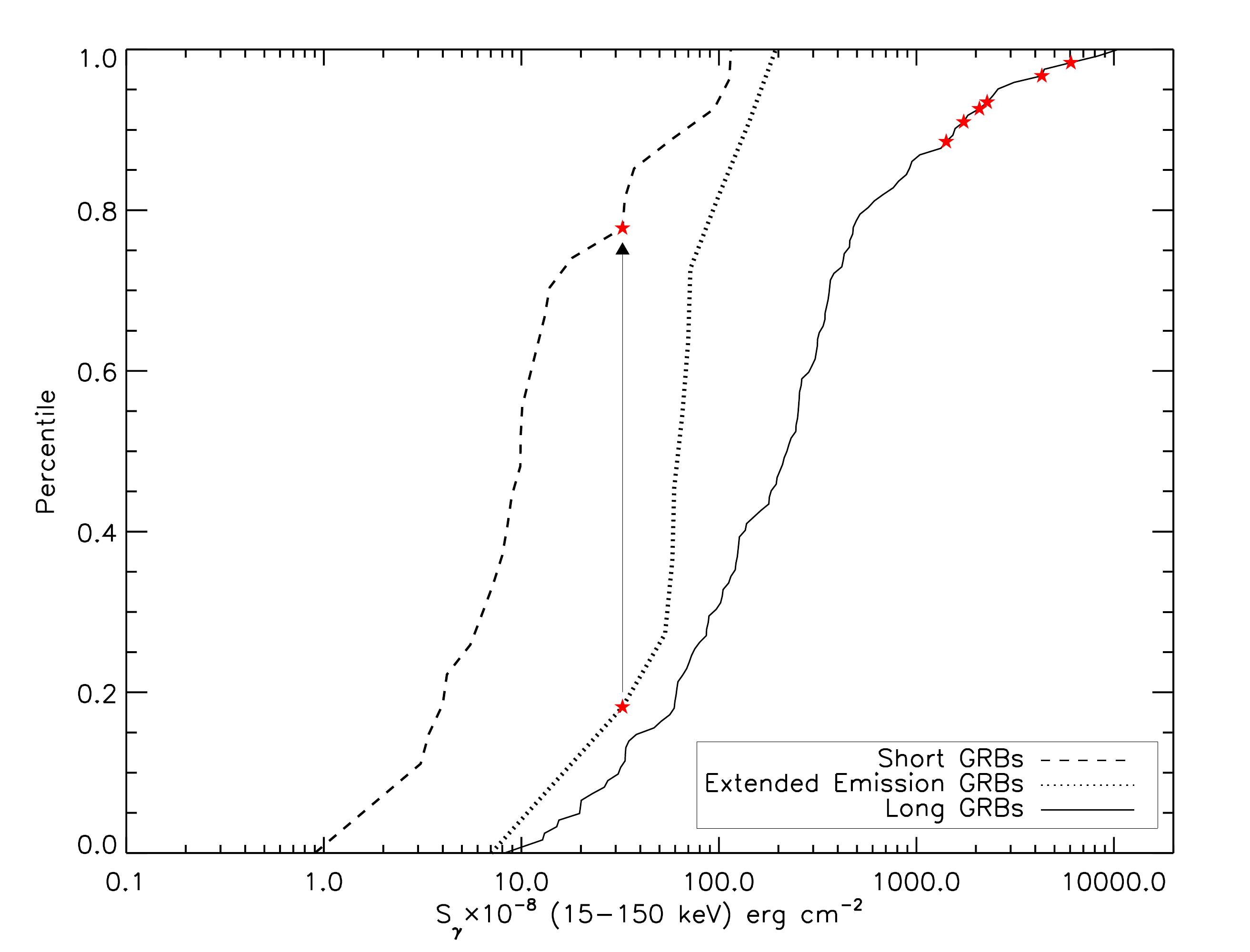}
\caption{Distribution curve for 343 BAT bursts from April, 2005 to June, 2009, and 7 LAT bursts as a function of fluence.  The stars indicate the LAT-detected GRBs, also observed by \textit{Swift}, using the predicted BAT fluence.  GRB 090510A is shown on both the short and extended emission curves, joined by an arrow. }
\label{fig:BATPercentile}
\end{figure}

We compare the calculated fluences to a sample of 343 BAT-triggered bursts from April, 2005 to June, 2009.  The sample is comprised of both short and long bursts, across a wide spread of energies.  The percentile ranking as a function of BAT fluence (or calculated fluence) is shown in Fig.~\ref{fig:BATPercentile}.   All but one of the LAT-detected bursts are brighter than 88\% of the BAT sample of bursts.  The exception is the only LAT short burst, GRB 090510A.

Ukwatta et~al. (2009) reported \textit{possible} soft, extended emission associated with GRB 090510A.  Because it was at a higher redshift than most short GRBs, $z$=0.903 (Rau et~al. 2009), BAT couldn't confirm any extended emission (De Pasquale et~al. 2010).   When we compare GRB 090510A to the BAT-triggered extended emission short GRBs, we find that it is only brighter than 18\% of the sample.  If extended emission is in fact present, GRB 090510A would be one of the lowest fluence extended emission bursts triggered by the BAT.  If there was no extended emission associated with GRB 090510A, then it would be brighter than $\sim$77\% of all BAT-triggered non-extended emission short bursts, making it a better corollary to the long LAT GRBs.

\begin{figure}
\includegraphics[scale=0.67]{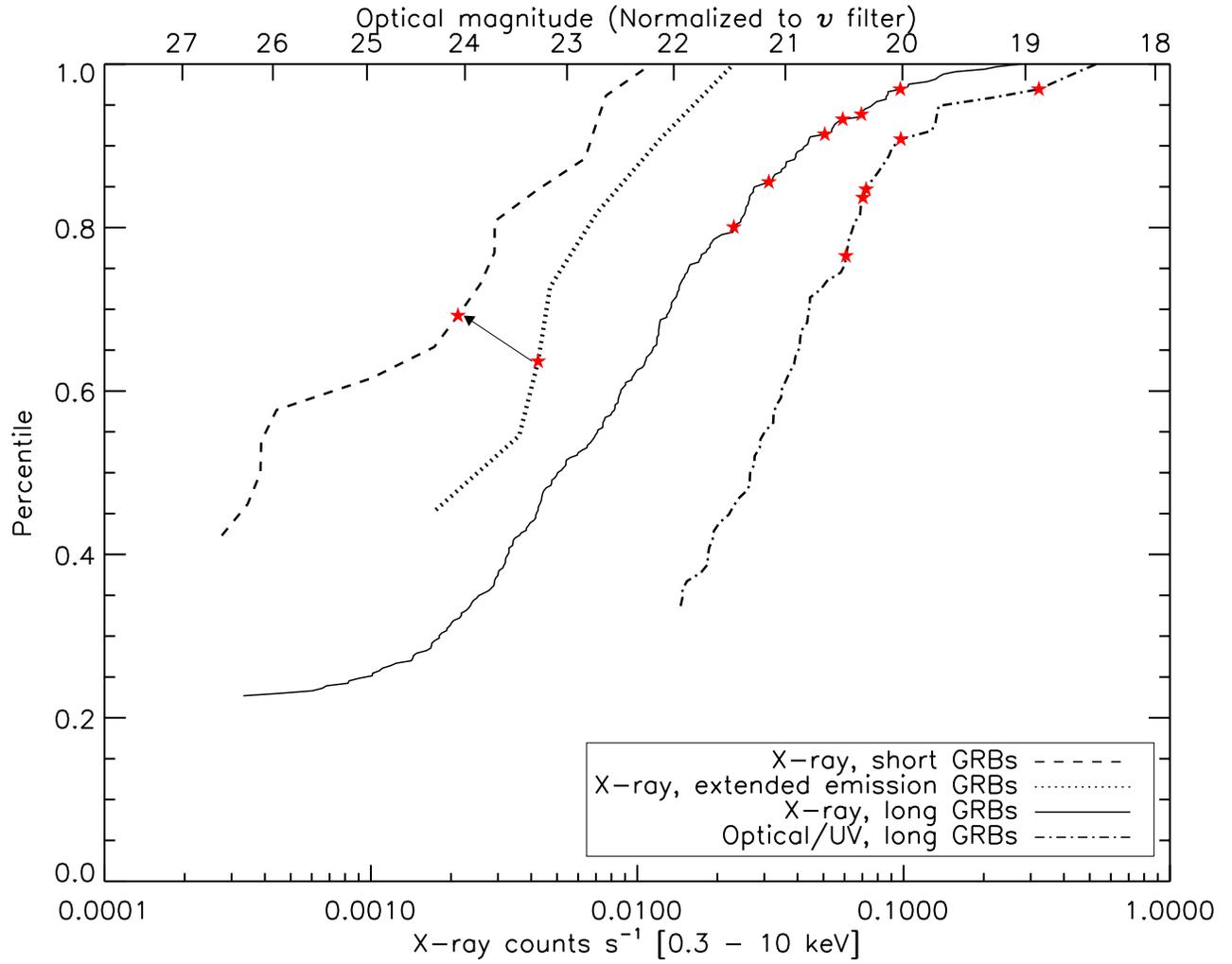}
\caption{X-ray and optical distribution curves.  X-ray curves using flux from 284 XRT afterglows.  Long bursts flux taken at 70 ks, short and extended emission at 35 ks.  Short burst curve is shifted to left by a factor of 2, for clarity.  GRB 090510A is shown on both the short and extended emission curves, joined by an arrow.  Optical distribution curve in both counts s$^{-1}$ and magnitudes in UVOT \textit{v} filter at 70 ks.  Observations resulting in upper limits are not included.  Stars indicate LAT bursts.}
\label{fig:XrayOpticalPercentile}
\end{figure}

We have shown that the long LAT-detected GRBs are brighter than 88\% of BAT-triggered bursts and that the lone short burst is also brighter than $\sim$77\% of other short bursts.  To test whether this trend continues to the X-ray and UV/optical wavelengths, we also compared the optical afterglows of the LAT sample to BAT-triggered bursts with XRT and UVOT afterglows.  We compared the X-ray flux of the LAT burst, in counts s$^{-1}$, at $\sim$70 ks to a selection of 314 X-ray light curves taken from the XRT light curve repository (Evans et~al. 2007, Evans et~al. 2009).  GRB 090510A was only detected by the XRT until $\sim$35 ks, so we used the flux at 35 ks for comparing the short and extended emission bursts.  We find the X-ray afterglows of long LAT-triggered bursts are brighter than those of 80\% of the BAT-triggered bursts, as shown in Fig.~\ref{fig:XrayOpticalPercentile}.  The X-ray afterglow of GRB 090510A is brighter than 64\% (69\%) of the BAT extended emission (short) bursts.

We compared the optical flux in counts s$^{-1}$ at 70 ks to 103 bursts with UVOT afterglows included in The Second \textit{Swift} Ultra-Violet/Optical Telescope GRB Afterglow Catalog (P. Roming et~al. 2010, in preparation).  All light curves were normalized to the \textit{v} filter and extrapolated to 70 ks (if necessary) for our comparison.  Our preliminary results, shown in Fig.~\ref{fig:XrayOpticalPercentile}, indicate that the optical afterglows of long LAT bursts are brighter than 77\% of BAT-triggered optical afterglows, with GRB 090926A falling in the top 3\% of optical afterglow brightness.  Additionally, GRB 090510A is one of only two extended emission GRBs, or one of five short GRBs, still detected by the UVOT at 70 ks.  Regardless of which category (short or extended emission) GRB 090510A belongs to, it is brighter than $\sim$90\% of other short/extended emission optical afterglows.

\section{Conclusions}

We have presented the \textit{Swift} and \textit{Fermi} observations of GRB 090926A, a recent LAT-detected GRB with a bright, long lived optical afterglow detected by UVOT.  We have compared this burst, to other LAT-detected and \textit{Swift} BAT bursts in an attempt to show whether the GRBs detected by the LAT are simply brighter than the average BAT-triggered GRB or whether they represent a new type of GRB that commonly exhibit bright, long duration optical afterglows due to some form of energy injection.

We find that the LAT-detected bursts are generally brighter than their BAT-triggered counterparts.  We find that their fluence is consistently higher than the `average' BAT burst, and that their X-ray and UV/optical afterglows are brighter than $\sim$80\% of BAT GRBs.

Although we are working with a small sample of LAT bursts, and therefore suffer the consequences of small number statistics, our preliminary results indicate that LAT bursts exhibit bright late time X-ray and UV/optical afterglows because they are brighter at all wavelengths than the `average' burst, assuming the higher than average fluence can be extrapolated down to X-ray and UV/optical wavelengths.  This seems to be the most likely explanation, given the known correlation between prompt emission and afterglow emission brightness (Gehrels et~al. 2008).  We cannot say definitively, however, that this is the reason for the bright afterglows at late times, due to the presence of flares, which indicate possible late time central engine activity that could cause a rebrightening.  Without coverage of the early afterglow, it is impossible to say how the afterglow arrived at the state in which we observe it $\sim$70 ks after the trigger.  If we simply extrapolate the optical light curve of GRB 090926A backward, we find that they could have peaked as high as \textit{v} = 10 mag within the first hundred seconds after the trigger.  Extrapolating the LAT spectrum of GRB 090926A to the \textit{v} band yields a peak magnitude of \textit{v} = 4, or if we assume a cooling break at GeV energies, the spectral index changes to $\beta \approx -0.76$, yielding a magnitude of \textit{v} = 15, consistent with our extrapolation backwards and the idea that LAT bursts are uniquely bright at all wavelengths.  However, if the early afterglow was fainter than \textit{v} $\approx$ 15 mag, then some sort of sustained energy injection would be required to keep the flux elevated at a level where we could then observe the bright afterglow at 70 ks after the trigger.  Such an energy injection would test our current theoretical understanding of GRB optical afterglows.  Our ability to determine the true nature of LAT-detected burst is contingent on our ability to follow-up LAT-detected GRBs at earlier times than has been achieved with the current sample.


\newpage

\begin{table}
\begin{center}
\caption{The 7 LAT observed bursts that have been observed by \textit{Swift} and detected by the XRT.  The first 6 columns give the burst parameters as measured by the \textit{Fermi} GBM  (Ghisellini et al. 2010), including those for GRB 090510A, which was also localized by \textit{Swift} BAT.  The last column gives the predicted BAT fluences as extrapolated from the GBM parameters.  The indices $\beta_{1 GBM}$ and $\beta_{2 GBM}$ are the low and high Band spectral parameters, respectively.  Fluences, S, are given in [ergs cm$^{-2}$].  We use a Band-function for the GBM spectrum, with the exception of GRB 090323, for which a cut off power-law model is adopted.  $^{\textit{a}}$: Actual fluence observed by BAT.\label{tab:ParamsFluence}}
\begin{tabular}{lllllll}
\tableline\tableline
GRB &  S$_{GBM}$  & T$_{90}$ & $\beta_{1 GBM}$ & $\beta_{2 GBM}$ & E$_{Peak}$ & S$_{BAT}$ \\
  &  8-10$^{4}$ keV     &   s      &       &       &  keV    &  15-150 keV        \\
\tableline
080916C      &  (1.6$\pm$0.2)$\times10^{-4}$     & 66             & -0.91$\pm$0.02    & -2.08$\pm$0.06  & 424$\pm$24 & 1.735$\times10^{-5}$ \\
090323        &  (1.32$\pm$0.03)$\times10^{-4}$  & $\sim$150 & -0.89$\pm$0.03   & . . .                      & 697$\pm$51 & 2.08$\times10^{-5}$ \\
090328        &  (1.52$\pm$0.02)$\times10^{-4}$  & $\sim$25  & -0.93$\pm$0.02    & -2.2$\pm$0.1      & 653$\pm$45 & 1.415$\times10^{-5}$\\
090510A      &  (2.3$\pm$0.2)$\times10^{-4}$     & 1               &  -0.80$\pm$0.03   & -2.6$\pm$0.3      & 4400$\pm$400 & 3.25$\times10^{-7} (5.57\times10^{-7})^\textit{a}$\\
090902B      &  (5.4$\pm$0.04)$\times10^{-4}$    & $\sim$21  &  -0.696$\pm$0.012&-3.85$\pm$0.25  & 775$\pm$11 & 6.05$\times10^{-5}$\\
090926A      &  (1.9$\pm$0.05)$\times10^{-4}$    & 20$\pm$2&  -0.75$\pm$0.01   & -2.59$\pm$0.05  & 314$\pm$4 &	4.316$\times10^{-5}$\\
091003        &  (4.16$\pm$0.03)$\times10^{-5}$  & 21$\pm$0.5& -1.13$\pm$0.01 & -2.64$\pm$0.24  & 86.2$\pm$23.6 & 2.279$\times10^{-5}$\\
\hline
\tableline
\end{tabular}
\end{center}
\end{table}

\begin{thebibliography}{99}

\bibitem[{{Amati} {et~al.} (2002)}]{} {Amati}, L., et~al. 2002, A\&A, 390, 81

\bibitem[{{Atwood} {et~al.} (2009)}]{} {Atwood}, W.~B., et~al. 2009, ApJ, 697, 1071

\bibitem[{{Barthelmy} {et~al.} (2005)}]{} {Barthelmy}, S.~D., Barbier, L.~M., Cummings, J.~R., et~al. 2005, SSRv, 120, 143

\bibitem[{{Bissaldi} {} (2009)}]{} {Bissaldi}, E. 2009, GCN 9933

\bibitem[{{Beuermann} {et~al.} (1999)}]{} {Beuermann}, K., Hessman, F.~V.,  Reinsch, K., et~al. 1999, A$\&$A, 352, 26

\bibitem[{{Burrows} {et~al.} (2005)}]{} {Burrows}, D.~N., Hill, J.~E., Nousek, J.~A., et~al. 2005a, SSRv, 120, 165

\bibitem[{{Burrows} {et~al.} (2005)}]{} {Burrows}, D.~N., et~al. 2005b, Science, 1833, 2005

\bibitem[{{Chincarini} {et~al.} (2007)}]{} {Chincarini}, G., et~al. 2007, ApJ, 671, 1903

\bibitem[De Pasquale et al.(2010)]{2010ApJ...709L.146D} De Pasquale, M., et~al.\ 2010, ApJL, 709, L146

\bibitem[{{Evans}{et~al.} (2007)}]{} {Evans}, P.~A., Beardmore, A.~P., Page, K.~L., et~al. 2007, A\&A, 469, 379

\bibitem[{{Evans}{et~al.} (2009)}]{} {Evans}, P.~A., et~al. 2009, MNRAS, 397, 1177

\bibitem[{{Gehrels} {et~al.} (2004)}]{} {Gehrels}, N., et~al. 2004, ApJ, 611, 1005

\bibitem[{{Gehrels} {et~al.} (2008)}]{} {Gehrels}, N., et~al. 2008, ApJ, 689, 1161

\bibitem[{{Ghisellini} {et~al.} (2010)}]{} {Ghisellini}, G., Ghirlanda, G., \& Nava, L. 2010, MNRAS, 403, 926

\bibitem[{{Gronwall} {et~al.} (2009)}]{} {Gronwall}, C., \& Vetere, L. 2009, GCN 9938

\bibitem[{{Kalberla} {et~al.} (2005)}]{} {Kalberla}, P.~M.~W., Burton, W.~M., Hartmann, D., Arnal, E.~M., Bajaja, E., Morras, R., \& P\"{o}ppel, W.G.L. 2005, A\&A, 440, 775

\bibitem[{{Krimm} {et~al.} (2007)}]{} {Krimm}, H.~A., et~al. 2007, ApJ, 665, 554

\bibitem[{{Krimm} {et~al.} (2009)}]{} {Krimm}, H.~A., et al. 2009, ApJ, 704, 1405

\bibitem[{{Malesani} {et~al.} (2009)}]{} {Malesani}, D., et~al. 2009, GCN 9942

\bibitem[{{Margutti} {et~al.} (2010)}]{} {Margutti}, R. et~al. 2010, \href{http://lanl.arxiv.org/abs/1004.1568v1}{arXiv:1004.1568v1}

\bibitem[{{Maxham and Zhang} (2009)}]{} {Maxham}, A. \& Zhang, B. 2009, \href{http://arxiv.org/abs/0911.0707}{arXiv:0911.0707v1}

\bibitem[{{Meegan}{et~al.} (2009)}]{} {Meegan}, C., et~al. 2009, ApJ, 702, 791

\bibitem[{{Oates}{et~al.} (2009)}]{} {Oates}, et~al. 2009, MNRAS, 395, 490

\bibitem[{{Rau}{et~al} (2009)}]{} {Rau}, A., et~al. 2009, GCN 9353

\bibitem[{{Roming}{et~al.} (2005)}]{} {Roming}, P.~W.~A., Kennedy, T.~E., Mason, K.~O., et~al. 2005, SSRv, 120, 95

\bibitem[{{Roming}{et~al.} (2009)}]{} {Roming}, P.~W.~A., et~al. 2009, ApJ, 690, 163

\bibitem[{{Uehara}{et~al.} (2009)}]{} {Uehara}, T., Takahashi, H., \& McEnery, J. 2009, GCN 9934

\bibitem[{{Ukwatta}{et~al.} (2009)}]{} {Ukwatta}, T.~N., et~al. 2009, GCN 9337

\bibitem[{{Wu}{et~al.} (2005)}]{} {Wu}, X.~F., Dai, Z.~G., Wang, X.~Y., Huang, Y.~F., Feng, L.~L., \& Lu, T. 2005, \href{http://arxiv.org/abs/astro-ph/0512555v1}{arXiv:astro-ph/0512555v1}

\bibitem[{{Yonetoku}{et~al.} (2004)}]{} {Yonetoku}, D., et~al. 2004, ApJ, 609, 935

\bibitem[{{Zhang}{et~al.} (2002)}]{} {Zhang}, B. \& M{\'e}sz{\'a}ros, P. 2002, ApJ, 581, 1236

\bibitem[{{Zhang}{et~al.} (2006)}]{} {Zhang}, B., et~al. 2006 ApJ, 624, 354

\end{thebibliography}
\end{document}